\documentclass[aps,prb,preprint,showpacs,a4paper,unsortedaddress,amsmath,amssymb,
byrevtex]{revtex4}
\usepackage{times}
\usepackage[]{graphicx}
\begin{document}



\title[Transport properties of a molecular junction]{Study of
transport properties of a molecular junction as a function of distance
between the leads }


\author{V\'{\i}ctor M. Garc\'{\i}a-Su\'arez$^1$}
\author{Tomasz Kostyrko$^2$}
\author{Steven Bailey$^1$}
\author{Colin Lambert$^1$}
\author{Bogdan R. Bu{\l}ka$^3$}

\affiliation{$^1$ Department of Physics, Lancaster University,
Lancaster, LA1 4YB, U. K.}

\affiliation{$^2$ Institute of Physics, Adam Mickiewicz
University, Umultowska 85, 61-614 Pozna{\'n}, Poland }

\affiliation{$^3$ Institute of Molecular Physics, Polish Academy
of Science, Smoluchowskiego 17, 60-179 Pozna{\'n}, Poland}

\begin{abstract}
We consider a model of a molecular junction made of BDT (benzene
dithiol) molecule trapped between two Au(100) leads. Using the ab
initio approach implemented in the SIESTA package we look for the
optimal configuration of the molecule as a function of a distance
between the leads.  We find that for the distance long enough the
energy of the system is minimized when the molecule is bonded
asymmetricaly, i.e. chemisorbed to one of the leads, whereas for the
distance shorter than 12~\AA\/ the energy is minimized for the molecule
sitting in the middle between the leads. We discuss possible
consequences of the above findings for the transport properties of the
junction.
\end{abstract}
\maketitle                   

Molecular junctions made of two leads bridged by a single molecule are
intensly studied recently both experimentally and theoretically as
basic units in the systems of molecular
electronics\cite{Reed-97,McCreery-04,Nitzan-03}.  Unlike the
traditional silicon based devices, the electronic transport through
the molecular devices very much depend on subtle details of molecular
configuration between the leads, due to importance of interference
effects at the nanoscale. In several {\it ab initio} works the
influence of a type of bonding between the metal surface and the
bridging atom of the molecule (on-top, hollow place, bridge contact
structures) on the current\---voltage (I-V) characteristics was
thoroughly studied for both Au(111) and Au(100)
surfaces\cite{Ke-05,Kondo-06,Grigoriev-06}. Moreover the variability
of the transport properties of the junctions with the change of width
of the leads (being also the kind of interference phenomenon) was
demonstrated\cite{Ke-05}.  So far however a full understanding of the
relation between the distance between the leads and the details of the
I-V characteristics seems to be absent in the literature.

%
%
In this paper we study the evolution of electronic and transport
properties of a molecular junction with change of the distance between
the leads. We consider benzene dithiol (C$_6$H$_4$S$_2$ or BDT)
molecule placed between the gold leads, with the sulfur atoms bound to
the Au(100) surface at the hollow positions, i.e. at the centers of
squares formed by surface Au atoms. We first look for the optimal
configuration of the molecule between the leads and analyze how this
configuration depends on the distance, and next we compute the
transmission function and the I-V dependence for the optimized
geometry of the junction.

\begin{figure}[h]
\centerline{\includegraphics[width=14cm,angle=0]{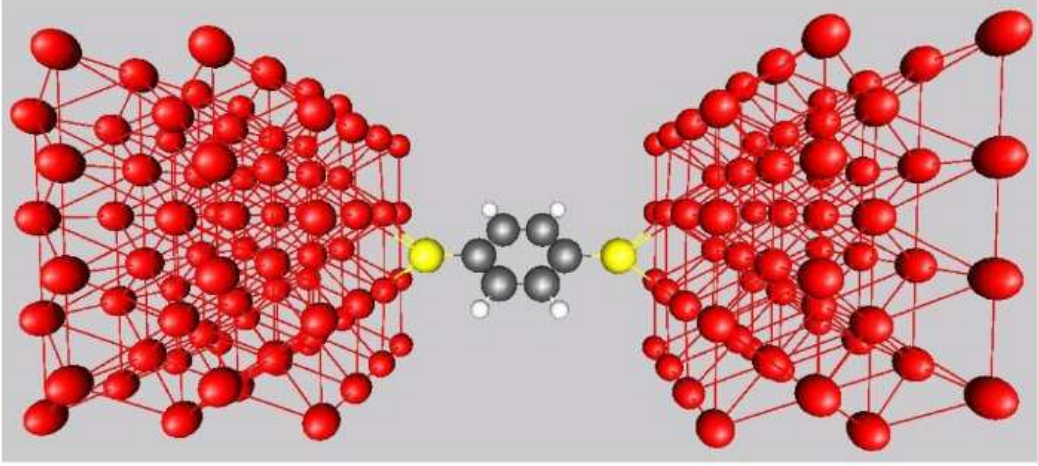}}
\caption{\label{junction} A perspective view of the optimized unit
cell of the Au(100)-BDT-Au(100) molecular junction obtained from
gOpenMol for $d_{lead}$=11~\AA. }
\end{figure}

%
%
To find the optimal configuration of the Au(100)-BDT-Au(100) junction
we apply the density functional approach as implemented in the SIESTA
package\cite{siesta}. In our SIESTA computation we consider a periodic
3D lattice with a unit cell consisting of BDT molecule and fragments
of the opposite leads, which include total of 144 Au atoms (see Figure
\ref{junction}).

Such a lattice represents well enough the typical experimental setup
as used in the mechanically controlled break junction (MCBJ)
experiment\cite{Reed-97} which involves a single BDT molecule provided
a number of molecular orbitals of BDT is small enough as compared to a
number of Au atoms. Moreover, we have to avoid overlapping between the
basis functions of BDT molecules from the neighbouring unit cells what
means that the transverse size of the unit cell should be greater than
the sum of the transverse size of the molecule and the maximal
diameter of the basis function.

All the numerical results presented below are obtained using local
density approximation (LDA) for the exchange\---correlation functional
of the DFT method\cite{Perdew-81}. The results of other works as well
as our own calculations show that neither nonlocal correction
(i.e. GGA) nor a spin\---dependent functional (LSDA) change the
results substantially.  We use the
Troullier\---Martins\cite{Troullier-91} pseudopotential to represent
potential of the atomic cores. Our function basis is restricted to
single\---zeta (SZ) functions and we checked some results using the
DZP basis. The radii of all the pseudoorbitals were determined by the
pseudo\---atomic orbital energy shift equal to 0.02~Ry. With the above
restrictions all our electronic structure and transport computations
(except the ones using the DZP basis) were possible using 2GB of a
computer memory.  We find the optimal configuration of the junction
with the conjugated gradient method, relaxing also the positions of 4
Au atoms binding the molecule in each lead. A locally stable
configuration is found numerically if the forces acting on each of the
relaxed atoms are less than 0.1~eV/\AA.

\begin{figure}[h]
\begin{center}
\begin{minipage}{14cm}
  \centerline{\includegraphics[height=7cm,angle=0]{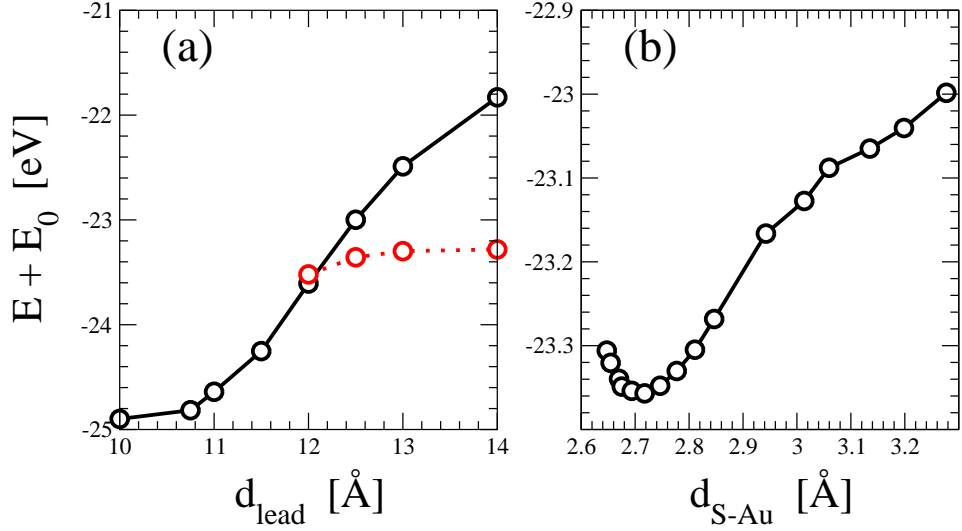}}
\end{minipage}
\end{center}
\caption{\label{distance_dependence} (a) Dependence of the energy
of the junction as a function of the distance between the Au
leads. Solid line: BDT molecule at the central position, dotted
line: BDT molecule shifted to one of the leads. For the shifted
molecule the distance between the S atom and the lead surface was
about~1.75~\AA. $E_0$=-13100~eV is a reference energy value. (b)
Energy of the stable configuration of BDT molecule as a function
of distance between the sulfur atom and the nearest Au atoms for
$d_{lead}=$12.5~\AA. The solution for $d_{S-Au}\sim$3.3~\AA\/
corresponds to the central position of the molecule. In the both
figures circles show locally stable configurations.}
\end{figure}

For a small distance between the leads, $d_{lead}\sim$10~\AA, we
find that the molecule centered at the middle between the leads
attains the local energy minimum. Computations with the molecule
shifted uniformly out of the center lead to stable configurations
with higher energy (or the molecule shifted back to the center by
the minimization procedure). The  energy increases steadily with
displacement from the central position. We thus conclude that for
a small distance between the leads the central position represents
the global minimum and there is no any other energetically stable
configuration. The increase of $d_{lead}$ reduces the overlap of
the S orbitals of BDT and Au orbitals from the leads surface and
the energy of the system increases (see
Figure~\ref{distance_dependence}). The central position remains
the global energy minimum until a critical distance
$d_{lead}\sim$12.25~\AA, where we find that the energy of the
system can be reduced by shifting the molecule towards either one
of the leads.

For the lead distance bigger than the critical one we find that
the global energy minimum is obtained for the binding sulfur atom
at about 1.75~\AA\/ from the nearest lead surface, corresponding
to the nearest S-Au distance about 2.7~\AA. The energy difference
between the central position and the displaced one rises with
increasing the distance between the leads and for a very large
$d_{lead}$ reaches the value $\sim$2~eV.

\begin{figure}[h]
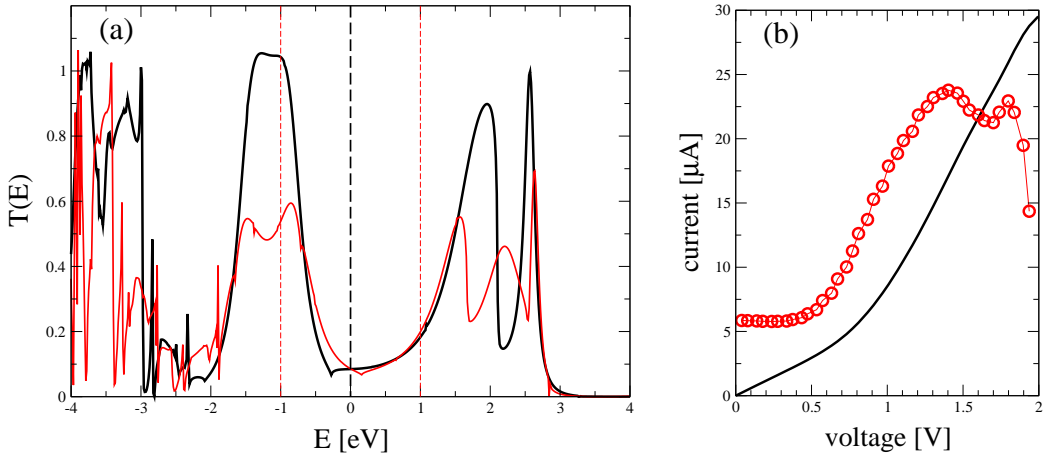

\begin{center}
\begin{minipage}{14cm}
    \includegraphics[height=6cm,angle=0]{transmission_10.75_1.eps}
    \hspace*{1em}
    \includegraphics[height=6cm,angle=0]{current_10.75_1.eps}
\end{minipage}
\end{center}
\caption{\label{transport} Transport properties of the junction
with $d_{lead}$=10.75~\AA. (a) Transmission as the function of
energy for $V$=0 (thick black curve) and $V$=2V (thin curve, red
in online edition). The broken lines show the corresponding
voltage windows, with $E=0$ being the position of the equilibrium
Fermi level. (b) Current as the function of the voltage (line) and
the voltage derivative of the current (circles).  }
\end{figure}

For the optimized junction configurations we computed transmission and
I-V characteristics using the SMEAGOL package\cite{smeagol} which is
an overlay on the SIESTA to allow computations of the transport in
nanostructures. The details of the applied method were described in a
recent paper by Rocha and coworkers. The current $J$ was computed
using the standard equation of the Landauer theory\cite{Nitzan-03}:
\begin{equation}
J = \frac{2e}{h}\,\int\,dE T(E,V) \left[ f_L(E)-f_R(E)\right]
\end {equation}
where $T(E)$ denotes the transmission function and $f_L$, $f_R$
are the Fermi functions corresponding to the left and the right
leads. The transmission function is computed using non-equilibrium
Green function method with the quasiparticle LDA Hamiltonian. For
the small distance between the leads ($d_{lead}$=10.75~\AA) energy
dependence of $T(E)$ exhibits a pseudogap region in the vicinity
of the equilibrium Fermi level and initially the current shows a
relatively slow linear increase with the voltage (see
Figure~\ref{transport}). With increase of the voltage the
source\---drain voltage window extends first over the region of
the HOMO level of the BDT, and next also the LUMO level, giving
rise to a steeper increase of the current and peaks in the
differential conductance.

The overall behaviour of the transmission and the current are
quite similar to the results of the papers of Xue and
Ratner\cite{Xue-03}, as well as the Ke {\it et al.}\cite{Ke-05}.
As in the cited papers, the current at $V$=2V is greater than the
experimental value by at least order of magnitude which may be due
to too poor treatment of electron correlations by the LDA method.
On the other hand, the equilibrium conductance obtained here is
several times smaller than the one computed in some other recent
papers(see e.g. \cite{Kondo-06}) and also the shape of our
differential conductance is more similar to the experimental
one\cite{Reed-97}.  We speculate that the difference may be due to
the greater transverse size of the Au leads in our paper.
%
%

\begin{figure}[h]
\begin{center}
\begin{minipage}{14cm}
    \includegraphics[height=6cm,angle=0]{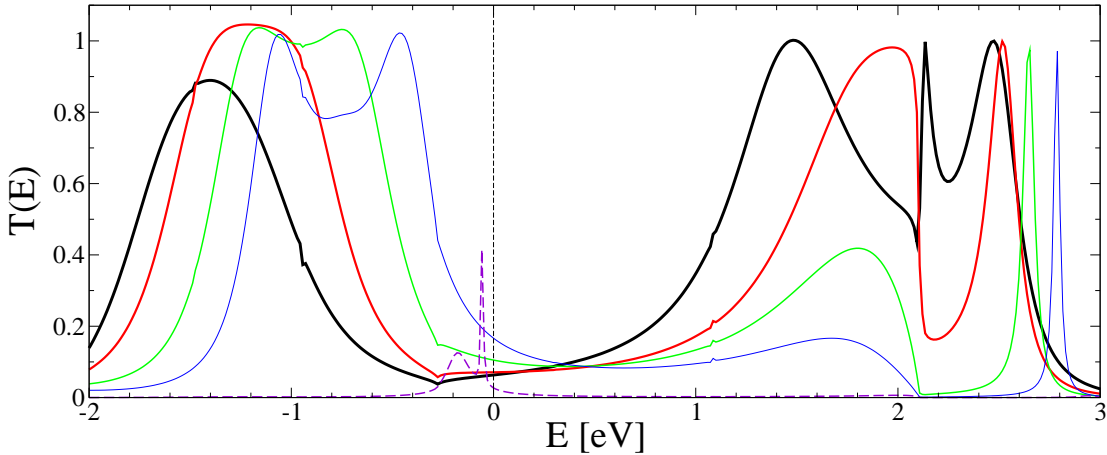}
\end{minipage}
\end{center}
\caption{\label{transport_all} Transmission as the function of
energy for $V$=0 for variable distance between the leads:
$d_{lead}$=10~\AA, 10.75~\AA, 11.5~\AA, and 12~\AA\/ (the line
thickness decreases with the distance). The broken line near $E=0$
shows the transmission for $d_{lead}$=12.5~\AA, with the BDT
molecule attached asymmetrically to one of the leads in the lowest
energy configuration from the Figure~\ref{distance_dependence}(b).
}
\end{figure}

In Figure~\ref{transport_all} we show the equlibrium transmission
functions in the lowest energy configurations for several values
of the distance between the leads. Until $d_{lead}\sim$12.5~\AA
the transmission changes gradually with the distance. For a small
$d_{lead}$ the maximum of $T(E)$ at the HOMO peak is smaller than
unity ($T\sim$0.9) and separated from $E_F$ by about $\sim$1.5~eV.
With increase of $d_{lead}$, the peak shifts towards $E_F$, and
finally it splits into two peaks, where T(E) is very close to
unity. This behaviour resembles the corresponding one of a simple
tight\---binding model of a two\---atom molecule. There, for a
coupling between the molecule and the leads stronger than some
critical value (depending on an interatom hopping), the two peak
structure merges into a single peak, which decreases with further
increase of the coupling\cite{Kostyrko-02}. This suggests that the
transmission of BDT can be partly understood in terms of the
two\---atom model, in which the internal part of BDT (all the
atoms except the sulfur ones) can be approximately represented by
a small and weakly energy dependent hopping between the external
atoms binding the molecule to the leads.

As a result of the upward shift of the HOMO peak, the transmission
at $E_F$ steadily increases with the distance between the leads
from $T(E_F)\sim0.05$ at $d_{lead}$=10~\AA\/ to $T(E_F)\sim0.15$
at $d_{lead}$=12~\AA. When the distance crosses the critical value
the central position of BDT molecule between the leads is no
longer energetically favoured. The transmission in the asymmetric
position is considerably smaller (see the broken curve in
Figure~\ref{transport_all}), since it is limited by a very weak
coupling with the more distant lead. As a  result, the current in
the asymmetric position is smaller than $\sim$0.2~$\mu$A for any
voltage.

In relation to an experiment the last results suggest, that with a
carefully controlled increase of the distance between the leads
one would observe first a steady rise of the transmission. At a
distance slightly bigger than 12.25~\AA\/ the current measured at
some constant voltage would suddenly drop. On the other hand, a
large value of current observed for a bigger distance would
suggest that a significant reorganisation of the junction takes
place. In fact on the basis of the molecular dynamical simulation
of Kr\"uger {\it et al.}\cite{Kruger-02} pulling of a string
of Au atoms by a molecule chemisorbed to an Au(111) lead can be
expected.

%
%

Finally the general comment is in order concerning the value of
the computed current, which greatly overestimates the ones
observed experimentally. So far this seems to be the common
deficiency of all the DFT and Hartree-Fock\cite{Shimazaki-06}
based approaches, at least for a weak coupling or relatively small
molecules. It remains to be seen if a full selfconsistent
implementation of a better treatment of the electron correlation
(e.g. SIC\cite{Toher-05}) will bring a substantial progress here.
Our results indicate also, that a better understanding of the
role of electron correlations in the transport through a simple
two\---atom molecule can be very helpful in this respect.

{\bf Acknowledgements}

This work is supported by the Ministry of Science and Higher Education
(Poland) and the project RTNNANO contract No. MRTN-CT-2003-504574, the
EPSRC, the DTI, the NWDA and the RCUK Basic Technology programme.

%
%

\end{document}